\documentclass[aps,prl,showpacs, twocolumn,superscriptaddress,reprint]{revtex4}

\usepackage{bm}
\usepackage{amssymb}
\usepackage{amsmath}
\usepackage{latexsym}
\usepackage{graphicx}
\usepackage{placeins}
\usepackage{booktabs}
\usepackage{subfigure}

\newcommand{\bvec}[1]{\mathbf{#1}}

\newcommand{\abs}[1]{\lvert#1\rvert}

\newcommand{\average}[1]{\left\langle#1\right\rangle}

\providecommand{\abs}[1]{\lvert#1\rvert}

\providecommand{\average}[1]{\left\langle#1\right\rangle}

\def\angstrom{\text{\AA}}

\def\vp{\bvec{p}}

\def\vr{\bvec{r}}

\def\vx{\bvec{x}}

\newcommand{\invcm}{\text{cm}^{-1}}

\begin{document}


\title{Momentum distribution, vibrational dynamics and the potential of mean force in ice} 
\affiliation{Program in Applied and Computational Mathematics, Princeton
University, Princeton, NJ 08544}
\affiliation{Department of Chemistry, Princeton University, Princeton, NJ 08544}
\affiliation{Department of Chemistry, Princeton University, Princeton, NJ 08544}
\affiliation{Department of Physics, Princeton University, Princeton, NJ 08544}
\affiliation{Computational Science, Department of Chemistry and Applied
Biosciences, ETH Zurich, USI Campus, Via Giuseppe Buffi 12, CH-6900 Lugano, Switzerland}
\author{Lin Lin}
\affiliation{Program in Applied and Computational Mathematics, Princeton
University, Princeton, NJ 08544}
\author{Joseph A. Morrone}
\altaffiliation{Present address: Department of Chemistry, Columbia University, New York NY 10027}
\affiliation{Department of Chemistry, Princeton University, Princeton, NJ 08544}
\author{Roberto Car}
\email{rcar@princeton.edu}
\affiliation{Program in Applied and Computational Mathematics, Princeton
University, Princeton, NJ 08544}
\affiliation{Department of Chemistry, Princeton University, Princeton, NJ 08544}
\affiliation{Department of Physics, Princeton University, Princeton, NJ 08544}
\author{Michele Parrinello}
\affiliation{Computational Science, Department of Chemistry and Applied
Biosciences, ETH Zurich, USI Campus, Via Giuseppe Buffi 12, CH-6900 Lugano, Switzerland}


\begin{abstract} 
  
  By analyzing the momentum distribution obtained 
  from path integral and phonon calculations
  we find that the protons in hexagonal ice 
  experience an anisotropic quasi-harmonic effective potential
  with three distinct principal frequencies that
  reflect molecular orientation. Due to the importance of anisotropy, 
  anharmonic features of the environment cannot be extracted
  from existing experimental distributions that involve the spherical average. 
  The full directional distribution is required, and we give
  a theoretical prediction for this quantity that could be verified in future experiments.  
  Within the quasi-harmonic context, anharmonicity in
  the ground state dynamics of the proton is substantial and has quantal origin,
  a finding that impacts the interpretation of several spectroscopies.

\end{abstract}

\maketitle

Investigating the impact of hydrogen (H) bonding on molecular properties is the focus of intense research, 
but even behavior as fundamental as
the equilibrium dynamics of the protons participating in H bonds remains poorly understood.
Proton dynamics is reflected in the momentum distribution probed by deep inelastic neutron 
scattering (DINS)~\cite{Andreani}. Recent DINS studies of H bonded systems 
have made striking observations, such as the presence of a secondary feature in the tail of the spherically averaged distribution 
in confined water~\cite{GarbuioAndreaniImbertiEtAl2007}, and estimates of a surprisingly large quantum kinetic energy 
of the proton in undercooled water~\cite{supercool}. The secondary feature was attributed to quantum tunneling
between the two wells of an anharmonic 1D potential~\cite{GarbuioAndreaniImbertiEtAl2007}.   
It is not clear, however, to what extent the dynamics of an interacting many body system can be reduced
to that of a single proton along a bond. For instance, it has been pointed out 
that anisotropy can mimic features of a spherical distribution
that one might associate to anharmonicity in a 1D model~\cite{Soper2009}, and yet so far there is no conclusive study of this issue. 
To interpret experiments in confined and undercooled water, 
the unknown details of the molecular structure are a severe source of
difficulty.  However, even
in the simpler case of ice Ih, it is not clear if the physics can be captured by simple 
model potentials, and how anharmonicity, anisotropy and structural disorder influence the momentum distribution.   

In order to tackle these issues we consider the open path integral Car-Parrinello molecular dynamics 
(PICPMD) data for ice Ih that yielded the accurate 
spherical momentum distribution reported
in a prior publication~\cite{morrone08}. In this prior study, no attempt was made to relate the distribution to the equilibrium dynamics of the proton or
to investigate the role of the environment in terms of a potential of mean force. 
In simulations this task is facilitated by access to the full 3D distribution, in contrast
to experiments on polycrystalline samples, where only the spherically averaged 
distribution could be measured~\cite{reiter,Andreani}. In addition,
crystalline symmetry allows the
use of harmonic analysis to quantify the relation between the momentum distribution and vibrational dynamics, 
thereby elucidating the role of anharmonicity and disorder on the proton ground state.

We find that anisotropy stemming from the molecular orientations in the crystal has a larger effect on  
the momentum distribution than anharmonicity. The latter is effectively
described within a quasi-harmonic model and is particularly important
in the stretching motions, corroborating pump-probe laser experiments 
on the excited state dynamics of ice and water~\cite{WoutersenEmmerichsNienhuysEtAl1998,BakkerNienhuys2002}.
This finding impacts the interpretation of infrared and x-ray spectroscopies, 
and regarding DINS experiments, the large effect of molecular anisotropy implies
that it is not possible to unambiguously attribute to anharmonicity features of the spherically 
averaged distribution.  Substantially more information, capable of disentangling anisotropy from anharmonicity,   
can be extracted from the directional distribution, for which we now present 
the first theoretical prediction for a realistic system. 

The simulations of Ref.~\onlinecite{morrone08} sampled the end-to-end distribution of the
open Feynman paths of the protons~\cite{morrone07}, i.e. 
$\widetilde{\nu}(\bvec{x}) =\frac{1}{N_p} \sum_{i}
\widetilde{\nu_{i}}(\bvec{x})$  where the sum runs over the $N_p$
protons in the cell and the vector $\vx$ points from one end of the path to the other.    
The momentum distribution $\nu(\bvec{p})$ is the Fourier transform of $\widetilde{\nu}(\bvec{x})$.
For each distribution $\widetilde{\nu_{i}}(\bvec{x})$ we compute
the correlation matrix $C_{i,\alpha\beta} = \average{x_{\alpha}
x_{\beta}}$.  Within the statistical errors of the simulation the
eigenvalues $\{\sigma_k^2\}_{k=1}^{3}$ of $C_{i}$ are the same for all the
protons, while the associated eigenvectors
$\{{\bvec{v}}_{i,k}\}_{k=1}^{3}$ are proton specific directions
related by crystalline symmetry to the
directions of the other protons. This suggests an anisotropic
Gaussian form for the end-to-end distribution: 
$\widetilde{\nu}_{i}(\bvec{x}) \propto \exp\left(-\frac12 \vx^{T} C_{i}^{-1} \vx \right)$.  
A quantile analysis reported in the supplement~\cite{suppmat} fully supports this hypothesis. Thus the momentum distribution
is $\nu_{i} (\bvec{p}) \propto \exp\left(-\frac{1}{2\hbar^2}\vp^T C_{i} \vp \right)$, implying that the 
corresponding potential of mean force has the \textit{effective} harmonic form $V(\bvec{r}) = \frac{M}{2}\vr^T A_{i} \vr$, where $M$ 
and $\bvec{r}$ denote the proton mass and position. 
$A_{i}$ has eigenvalues $\omega^2_{k}$ and shares with $C_{i}$ the eigenvectors,
${\bvec{v}}_{i,k}$. The $\omega_k$ are related
to the $\sigma_{k}^2$ by,
\begin{equation}
  \frac{1}{\sigma_{k}^2} =
  \frac{M \omega_k}{2 \hbar}\coth\frac{\hbar\omega_k}{2k_BT},
  \label{eqn:sigmaomega}
\end{equation}
and $\omega_k$ and
${\bvec{v}}_{i,k}$ are denoted 
the principal frequencies and directions
of proton $i$. Since the principal frequencies do not depend on $i$ all the protons
have equivalent local environments within the simulation error bars. 

\begin{figure}[h]
  \begin{center}
    \includegraphics[width=0.42\textwidth]{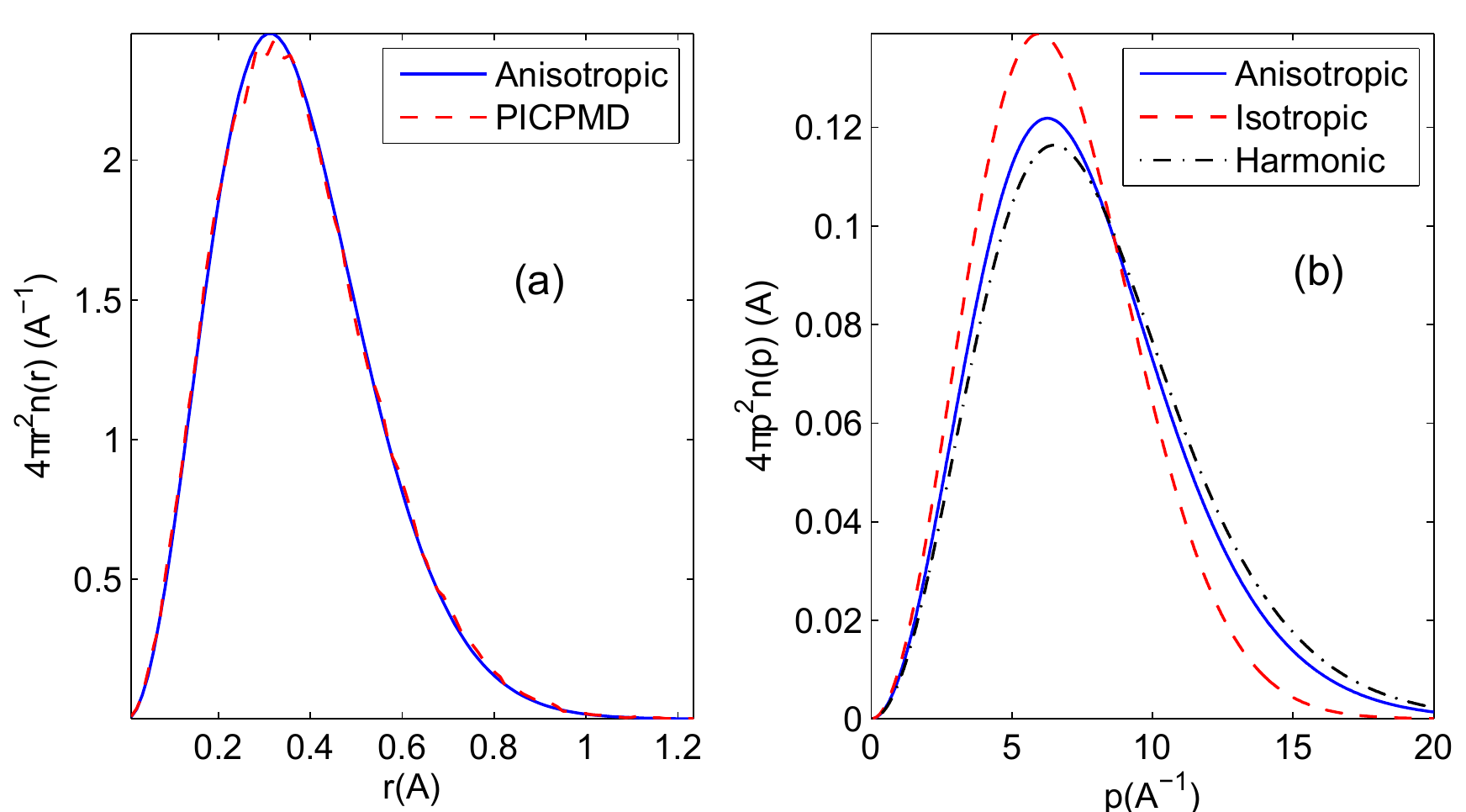}
    \caption{(color online) (a) The spherical end-to-end distribution
    directly collected from PICPMD data (red dashed line) compared with that reconstructed by the anisotropic
    fit (blue line). (b) Comparison of the spherical momentum distribution of the harmonic crystal (black dot-dashed line) with anisotropic (blue line) and isotropic (red dashed line) fits. }
    \label{fig:nrcomp}
  \end{center}
\end{figure}

By averaging over the protons we obtain the frequencies $\bar{\omega}_k$
with error bars in the first row of Table~\ref{tab:allfreq}. In
terms of the $\bar{\sigma}_{k}^2$
the spherically averaged end-to-end distribution takes the form, 
\begin{equation}
n(x)=\frac{1}{\sqrt{8 \pi^3} \bar{\sigma}_1 \bar{\sigma}_2
\bar{\sigma}_3}\int_{\abs{\bvec{x}}=x}\mathrm{d}\Omega~  e^{-
\frac{x_1^2}{2\bar{\sigma}_1^2}-\frac{x_2^2}{2\bar{\sigma}_2^2}-\frac{x_3^2}{2\bar{\sigma}_3^2}}. 
\label{eqn:sphericaldist}
\end{equation}
Fig.~\ref{fig:nrcomp}(a) shows that this curve differs negligibly from
the corresponding ``raw'' distribution extracted from the simulation, indicating that an effective harmonic model 
faithfully represents the spherically averaged data. 
Consistent with chemical intuition, the associated principal directions reflect the orientation of each water molecule 
in the crystal.  The principal axes corresponding 
to the highest frequency are close to the oxygen-oxygen nearest neighbor
directions, whereas the eigenvectors associated with the
middle and lowest frequency correspond respectively to directions in and
perpendicular to the HOH molecular plane. 

The PICPMD principal frequencies differ from their harmonic counterparts 
(see Table~\ref{tab:allfreq}). The latter  
were obtained with the phonon calculation discussed below. Thus the model that better represents 
the data is anisotropic and quasi-harmonic. We can now resolve, in the case of ice, a 
major issue that troubled the interpretation of experiments~\cite{Soper2009} by quantifying the relative importance of anisotropy and 
anharmonicity. We depict in Fig.~\ref{fig:nrcomp} (b)  the spherical distributions corresponding to, respectively, 
the quasi-harmonic model (first row of Table~\ref{tab:allfreq}),  the harmonic model (second row of Table~\ref{tab:allfreq}), 
and the isotropic model with frequency $\bar{\omega}=1186~\invcm$ that best fits the data. Anisotropy and 
anharmonicity are both significant, but anisotropy clearly has the larger effect. 
The isotropic model corresponds to a classical Maxwell-Boltzmann distribution with an effective temperature 
$\widetilde{T}=869 K$. In spite of $\widetilde{T}$ being significantly higher than the equilibrium temperature of the 
simulation ($T=269 K$), the isotropic model severely underestimates quantum effects, a finding that is also illustrated by a kinetic 
energy ($E_K = 111 \text{meV}$) approximately 30 percent smaller than
the simulation value ($E_K = 143 \text{meV}$).            

\begin{table}[h]
  \centering
  \begin{tabular}{p{1.5cm}|c|c|c|c}
    \hline
    & $\bar{\omega}_{1}(\invcm)$ & $\bar{\omega}_{2}(\invcm)$ & $\bar{\omega}_{3}(\invcm)$
    & $E_K(\text{meV)}$\\
    \hline
    PICPMD &  $2639 \pm 60$ & $1164 \pm 25 $ & $775 \pm 20$ & $143 \pm 2$\\
    \hline
    Harmonic & $3017.6\pm 8.2$ & $1172.5\pm 8.9$ & $870.3 \pm 14.6$ & $157.5\pm 0.3$\\
    \hline
  \end{tabular}
  \caption{Average proton principal frequencies and kinetic energies obtained from
  PICPMD and phonon calculations. The error bars reflect 
  statistical errors and physical effect of disorder in the PICMD and phonon data, respectively.}
  \label{tab:allfreq}
\end{table}

\begin{figure}[h]
  \begin{center}
    \includegraphics[width=0.22\textwidth]{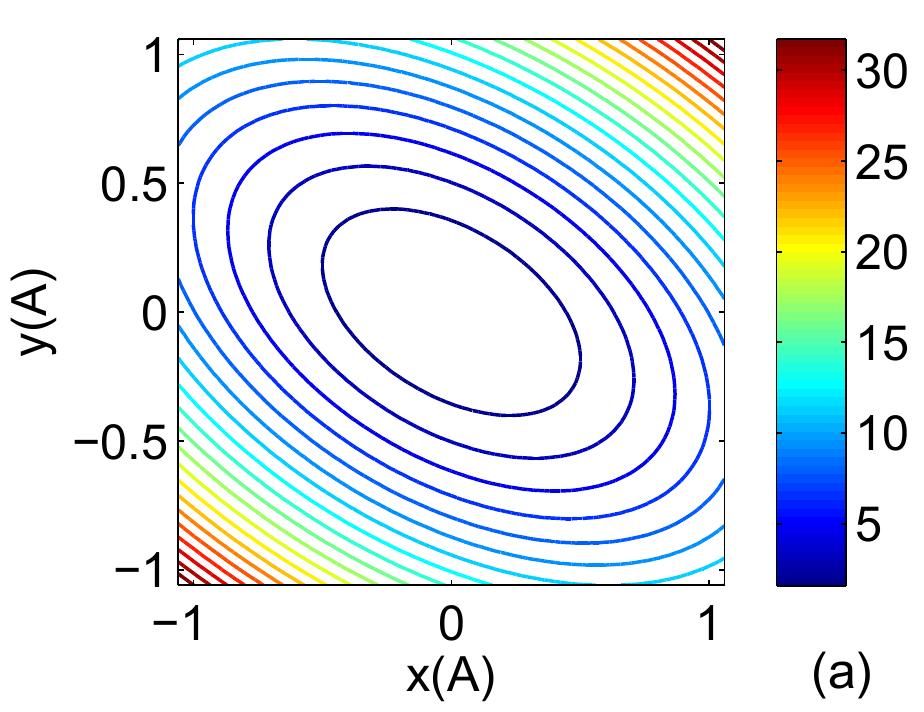}
    \includegraphics[width=0.21\textwidth]{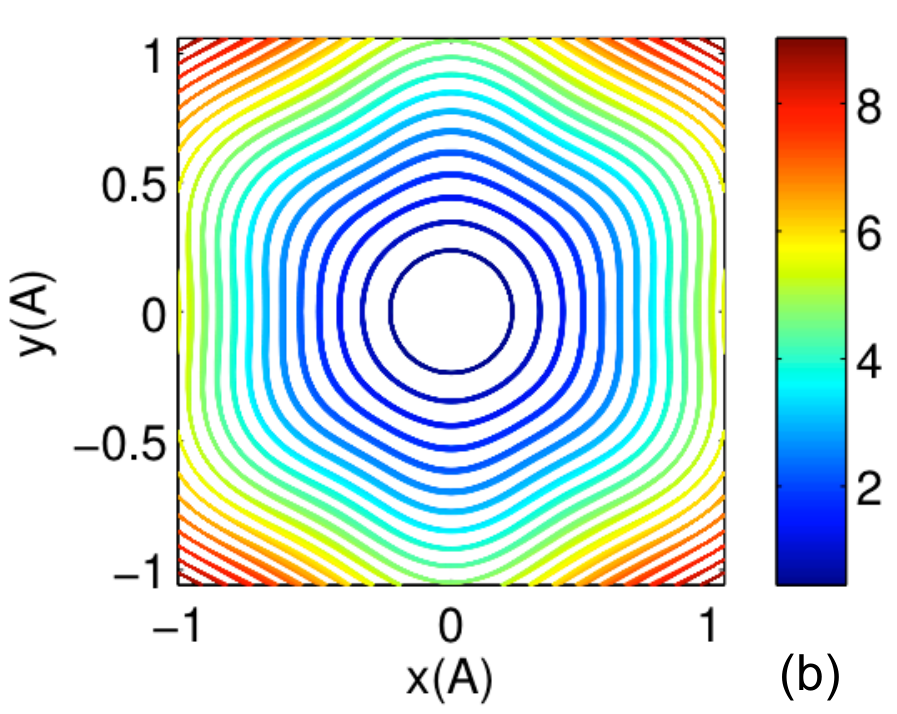}
  \end{center} 
  \caption{(color online) (a) ``Environmental part'' of the end-to-end
  distribution corresponding to one individual proton projected in the basal plane of ice Ih plotted in
  logarithmic scale. (b) ``Environmental part'' of the end-to-end
  distribution corresponding to the superposition of all protons
  projected in the basal plane of ice Ih plotted in logarithmic scale. 
  The super positioned end-to-end distribution reflects the symmetry of the oxygen
  sub-lattice.  The blurring of the contour lines reflects the
  disorder effect detected in the phonon calculation.  }
    \label{fig:lognxavg} 
\end{figure}

All the principal frequencies in Table~\ref{tab:allfreq} are well 
in excess of the equilibrium temperature, 
indicating that the dynamics of the proton is dominated by quantum zero-point motion.
Dependence of the molecular 
orientations upon the crystalline framework originates anisotropies that reflect the symmetry
of the environment in the momentum and end-to-end distributions. To study these effects
we focus on the latter distribution, which   
factorizes into the product of a spherical free-particle contribution and an anisotropic 
environmental component $\widetilde{n}_{V}$, i.e.
$\widetilde{n}(\bvec{x})\propto e^{-\frac{Mk_BT\bvec{x}^{2}}{2 \hbar^{2} } }\widetilde{n}_{V}(\bvec{x})$~\cite{lin10}.
Rather than extracting $\widetilde{n}_{V}(\bvec{x})$ directly from the PICPMD data, which would
be affected by substantial noise, we reconstruct $\widetilde{n}_{V}(\bvec{x})$ from the superposition of the 
individual proton contributions within the quasi-harmonic 
model.  Here we use the fact
that there are 24 unique orientations of the molecules in the hexagonal ice 
crystal~\cite{HaywardReimers1997}, and we also include the effects
of proton disorder estimated below in the phonon calculation. 
Fig.~\ref{fig:lognxavg} (a) depicts the log scale plot of one individual environmental
end-to-end distribution projected on
the basal plane of ice Ih. The elliptic shape of the contour comes
directly from the quasi-harmonic model.  Fig.~\ref{fig:lognxavg} (b)
illustrates the log scale plot of the superposition of all the environmental
end-to-end distributions.
The hexagonal shape of superpositioned distribution
is a striking manifestation of quantum mechanics as in classical
physics $\widetilde{n}_{V}(\bvec{x})$ is equal to $1$.
While the distribution is
spherical at the center, hexagonal character emerges at intermediate
displacements and becomes pronounced in the tail of the
distribution where blurring of the contour lines
due to disorder can be detected.
Experiments on ice Ih have only measured the spherical
distribution~\cite{reiter}  but it is likely that
the full three dimensional distribution should become
accessible in the future with improved instrumentation
and preparation techniques. Directional momentum distributions have already 
been reported for materials such as KDP \cite{kdp} and
Rb$_{3}$H(SO$_4$)$_2$~\cite{homouz}.  It should be noted, however,
that the greatest sensitivity to anisotropy is in the exponential tail of the distribution, a
finding indicating that substantial resolution may be necessary to experimentally
disentangle anisotropy, anharmonicity and other environmental effects.

\begin{figure}[h]
\begin{center}
\includegraphics[width=0.28\textwidth]{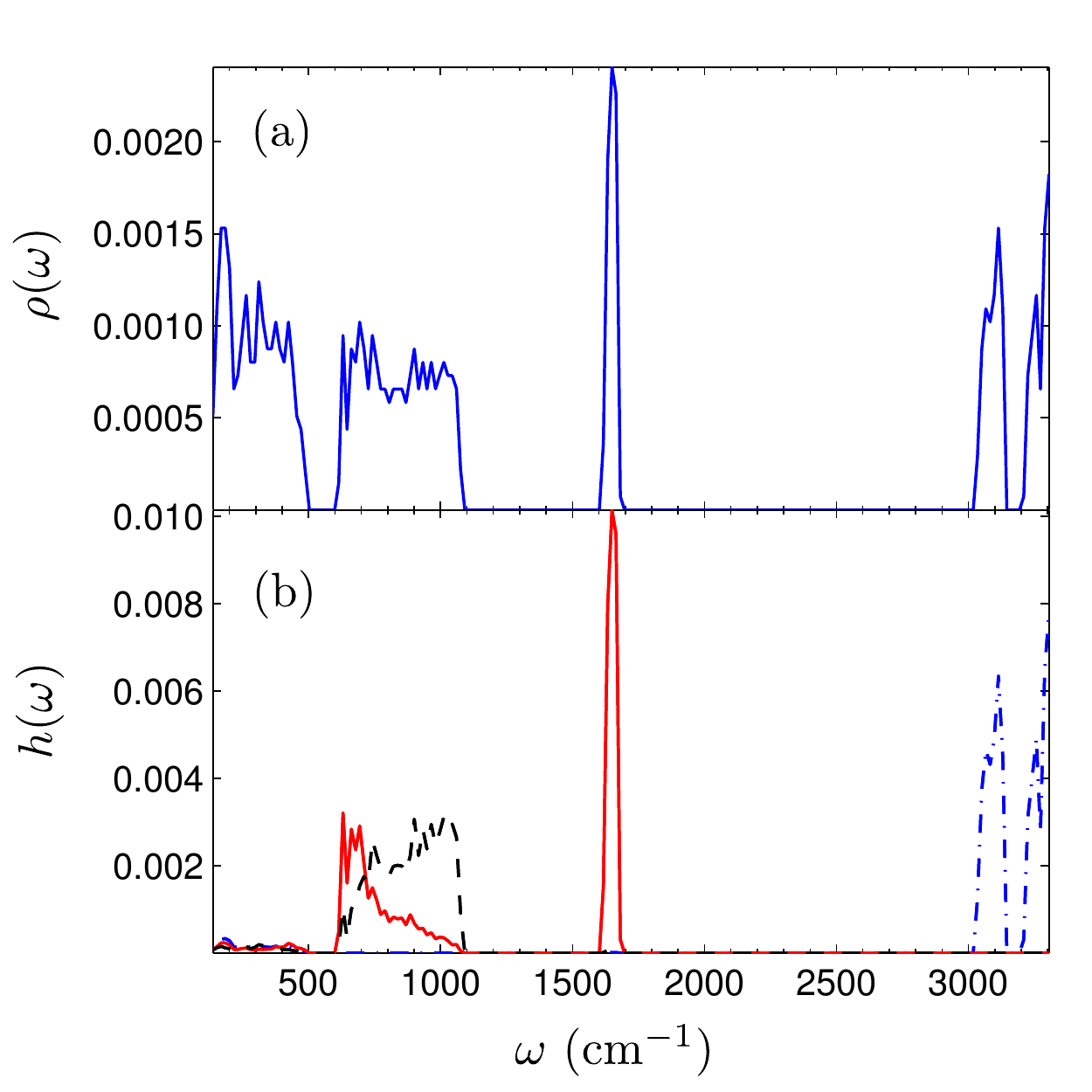}
\caption{(color online) (a) Density of states of the phonon spectrum.
(b) The population function for the
principal axes corresponding to 
$\bar{\omega}_{1}$ (blue dot-dashed line), 
$\bar{\omega}_{2}$ (red solid line) and $\bar{\omega}_{3}$ (black dashed
line). Network modes below $500\invcm$ contribute
non-negligibly to all principal frequencies.}\label{fig:dos}
\end{center}
\end{figure}

Lastly, we discuss the relationship between the principal frequencies and the vibrational spectrum. 
 The latter includes four main features experimentally: a stretching band centered
at $\approx 3250$ $\invcm$ ~\cite{BertieWhalley1964}, a bending band centered at 
$\approx 1650$ $\invcm$  ~\cite{TaylorWhalley1964}, a wide librational band between 
$\approx 400\invcm$ and $1050\invcm$ ~\cite{BertieWhalley1964,PraskBoutinYip1968} and a band
of network modes below $\approx 400 \invcm$ ~\cite{Li1996}. These features are reproduced in the phonon
spectrum of ice that we calculate by diagonalizing the dynamical matrix. 
This calculation is performed with Qbox~\cite{Gygi2008} by
adopting the same supercell, electronic structure parameters and disordered proton 
configuration of the PICPMD simulation~\cite{morrone08}.  The dynamical matrix is calculated with a finite difference
method (grid size of $0.0053\angstrom$). The resulting phonon density of states shown in Fig.~\ref{fig:dos}
(a), agrees with experiment, and is consistent with previous calculations~\cite{MorrisonJenkins1999}, which 
did not include proton disorder, indicating that such effects have a small influence on the spectrum.
We indicate phonon frequencies and eigenvectors by $\omega_{k}^{ph}$ 
and $e_{i\alpha,k}$, respectively, where $\alpha$ are Cartesian components, 
$i,k=1,\cdots,3N-3$, and $N$ is the number of supercell atoms. 
In the quantum harmonic approximation the momentum distribution of particle $i$ of mass $M_i$  
has the anisotropic Gaussian form $\nu_{i} (\vp_{i}) \propto \exp\left(-\frac{1}{2}\vp_{i}^T {C_i^{ph}}^{-1}
\vp_{i}\right)$ with correlation matrix~\cite{Ceriotti2010_Lithium},
\begin{equation}
  \begin{split}
    C_{i,\alpha\beta}^{ph} = \average{p_{i,\alpha} p_{i,\beta}}
    = \sum_{k} e_{i\alpha,k} e_{i\beta,k}
    \frac{M_i \hbar\omega_k^{ph}}{2}\coth\left( \frac{\hbar\omega_{k}^{ph}}{2k_B
    T} \right).
  \end{split}
  \label{eqn:corrharmonic}
\end{equation}

As a consequence of disorder the eigenvalues of $C_{i,\alpha\beta}^{ph}$,   
depend on the proton index $i$. The harmonic average frequencies are reported 
in the second row of Table~\ref{tab:allfreq}.
The corresponding standard deviations originate almost entirely from ice disorder, being
at least an order of magnitude larger
than the numerical errors estimated from
the small asymmetry of the calculated dynamical matrix.
The statistical errors in the PICPMD simulation 
(Table~\ref{tab:allfreq}) are on average a few times larger than the harmonic estimate of 
disorder, confirming that, within error bars, all proton environments  
are equivalent. 
We expect that longer runs combined with better estimators of the end-to-end distribution~\cite{lin10}
should improve the statistical accuracy to the point that disorder effects
could become measurable in future simulations.    

The population function,
\begin{equation}
  h(\omega_k^{ph}; l) = \frac{1}{N_p} \sum_{i=1}^{N_p} \left( \sum_{\alpha=1}^{3} v_{i\alpha,l} e_{i\alpha,k}
  \right)^2,
  \label{eqn:population}
\end{equation}
gives the weight of the phonon  $k$ in the principal direction $l$ and is depicted in Fig.~\ref{fig:dos} (b). 
It is found that  $\bar{\omega}_{1}$ is $94\%$ stretching, $\bar{\omega}_{2}$ is $47\%$ bending  and
$48\%$ libration,
and $\bar{\omega}_{3}$ is $97\%$ libration. Taking only
stretching, bending, and libration into account, and using weights proportional to
 $h$ we infer that
$\bar{\omega}_{1}\sim 3160\invcm$, 
$\bar{\omega}_{2}\sim 1210\invcm$, and $\bar{\omega}_{3}\sim 895\invcm$.
In comparison, the values in the second line of Table~\ref{tab:allfreq} are red-shifted by contributions 
from network modes ($6\%, 4\%$, and $3\%$ to
$\bar{\omega}_{1},\bar{\omega}_2$, and $\bar{\omega}_3$, respectively), an intriguing 
result suggesting that fine details of the momentum distribution should reflect
intermediate range order properties of the H bond network.  
  
The potential energy surface is generated 
with the same protocol in path integral and phonon calculations.
We thus attribute the difference between the average principal
frequencies in the two rows of Table~\ref{tab:allfreq} to anharmonicity.
This originates from quantum delocalization, present in the PICPMD
simulation,  which causes the proton to sample the potential over an
extended range. Along the bond direction
the proton spans from $\approx -0.2\angstrom \text{ to} \approx +0.3\angstrom$ 
relative to the centroid of the path. This is much larger than the corresponding classical thermal spread
($\approx \pm 0.05\angstrom$) indicating that quantum anharmonicity is essentially
unaffected by temperature. 
The asymmetry of the quantal spread suggests that the first correction to the harmonic potential
depends cubically on displacement. 
A visualization of the approximate effective potential along the bond based on further calculations is given
in the supplement~\cite{suppmat} and shows that 
while cubic terms dominate in the ground state, higher order corrections become important at displacements larger than $\approx 0.3 \angstrom$.
In the supplement we also report harmonic estimates
of the quantum effects on oxygen, which are non-negligible albeit smaller than for the protons.

In conclusion, we find that to a large extent the momentum distribution in ice is
a simple anisotropic Gaussian distribution. This does not
mean, however, that ice behaves like a harmonic crystal as the principal
frequencies of the distribution differ from those of a harmonic crystal.
Anharmonicity, enhanced by H bonding, is appreciable in the libration dominated $\bar{\omega}_{3}$
and is particularly significant in the stretching dominated $\bar{\omega}_{1}$, 
in agreement with optical pump-probe 
experiments~\cite{BakkerNienhuys2002,WoutersenEmmerichsNienhuysEtAl1998}.
The quantal character of the anharmonicity is consistent 
with the observed T-independence of the lifetime of excited stretching modes 
in ice~\cite{WoutersenEmmerichsNienhuysEtAl1998}. Our findings have implications
for the calculation of observables in ice, such as infrared spectra, 
which typically ignore quantum
anharmonicity~\cite{ChenSharmaRestaEtAl2008}, 
and x-ray absorption spectra, which typically ignore quantum 
configurational disorder~\cite{ChenWuCar2010}.  
The approach presented here could be applied directly to the study of other crystalline H bonded systems,
and is also an important step towards a better understanding of the proton momentum 
distribution in disordered H bonded systems such as water under different conditions. 
In such cases only the spherically averaged momentum distribution is accessible in experiment and
simulation can provide essential microscopic
information to supplement and interpret the experimental data. 
Finally, we remark that while the qualitative picture emerging
from our calculations is robust, the path integral data have relatively
large error bars and the quantitative details depend on the
accuracy of the underlying Born Oppenheimer potential energy surface.
The latter should reflect  the known limitations of the GGA
functional used in this study~\cite{grossman10,santra2009} and 
comparisons with future high resolution experiments should help to clarify this issue.

This work is partially supported by NSF CHE-0956500 and
DOE DE-SC0005180 (LL and RC), and by ERC-2009-AdG-247075 (MP). We thank F.
Gygi for the help using Qbox. The calculations were performed
at the TACC under the NSF TeraGrid program, and on 
the TIGRESS computers at Princeton University. 


\end{document}